\documentclass[a4paper]{article}
\usepackage{url}
\usepackage{INTERSPEECH2020}
\usepackage{lipsum}
\usepackage{stfloats}
\usepackage{amssymb}
\usepackage{dutchcal}
\usepackage{diagbox}
\usepackage{pifont}
\usepackage{float}
\usepackage{multirow}
\usepackage{hhline} 

\newcommand{\xmark}{\ding{55}}
\newcommand{\cmark}{\ding{51}}

\title{Investigation of Data Augmentation Techniques for \\ Disordered Speech Recognition}

\name{Mengzhe Geng$^{1*}$, Xurong Xie$^{1,2*}$, Shansong Liu$^1$, Jianwei Yu$^1$,  Shoukang Hu$^1$, \\Xunying Liu$^1$, Helen Meng$^1$ }

\address{
  $^1$The Chinese University of Hong Kong\\
  $^2$Shenzhen Institutes of Advanced Technology, Chinese Academy of Sciences}
  
\email{\{mzgeng,ssliu,jwyu,skhu,xyliu,hmmeng\}@se.cuhk.edu.hk, xrxie@ee.cuhk.edu.hk}

\begin{document}

\maketitle
\begin{abstract}
Disordered speech recognition is a highly challenging task. The underlying neuro-motor conditions of people with speech disorders, often compounded with co-occurring physical disabilities, lead to the difficulty in collecting large quantities of speech required for system development. This paper investigates a set of data augmentation techniques for disordered speech recognition, including vocal tract length perturbation (VTLP), tempo perturbation and speed perturbation. Both normal and disordered speech were exploited in the augmentation process. Variability among impaired speakers in both the original and augmented data was modeled using learning hidden unit contributions (LHUC) based speaker adaptive training. The final speaker adapted system constructed using the UASpeech corpus and the best augmentation approach based on speed perturbation produced up to 2.92\% absolute (9.3\% relative) word error rate (WER) reduction over the baseline system without data augmentation, and gave an overall WER of 26.37\% on the test set containing 16 dysarthric speakers.
\end{abstract}

\let\thefootnote\relax\footnotetext{*Equal contribution was made between the first two authors.}

\section{Introduction}

Speech disorders affect millions of people all over the world, severely degrading their quality of life. They may be caused by a range of medical conditions, such as cerebral palsy \cite{whitehill2000speech}, Parkinson disease \cite{scott1983speech} and stroke \cite{jerntorp1992stroke}. Speech disorders are always associated with difficulties in controlling articulators responsible for speech, which creates a large mismatch against normal speech \cite{rosen2006parametric}. This makes 
the recognition of disordered speech a challenging task \cite{hux2000accuracy,young2010difficulties}. Meanwhile, as speech disorders always come together with loss of abilities in physical motion, there is a popular demand of hands-free and speech-enabled assistive technologies to help such people \cite{fried1985voice,hawley2012voice}. 

Resurgence of deep learning technologies in the past decade improved the performance of state-of-the-art automatic speech recognition systems \cite{xiong2018microsoft,saon2017english,thomas2019english,manohar2017jhu,luscher2019rwth,Park2019}. However, these systems are not directly usable by people with speech disorders due to the large mismatch against disordered speech. Hence, there has been growing research interest in developing deep neural network (DNN) based modeling techniques suitable for disordered speech \cite{christensen2012comparative,christensen2013combining,sehgal2015model,yu2018development,liu2019exploiting,liu2019use,hu2019cuhk}.
The underlying neuro-motor conditions of people suffering from speech impairment, often compounded with co-occurring physical disabilities, lead to the difficulty in collecting large quantities of disordered speech required for automatic speech recognition (ASR) system development \cite{kim2008dysarthric,choi2011design,rudzicz2012torgo}. In addition, varying speech disorder characteristics and levels of severity create a large variation among different dysarthric speakers. These factors combined impose limitations on the performance of current disordered speech recognition systems \cite{sehgal2015model, yu2018development,liu2019exploiting,liu2019use,hu2019cuhk}. 

Data augmentation proves to be an effective method for dealing with data sparsity and improving the performance of DNN acoustic models for normal speech recognition \cite{Park2019,bell2012transcription,jaitly2013vocal,kanda2013elastic,cui2015data,ko2015audio,hayashi2018back}, as it enables the neural networks to improve coverage and generalization. A wide range of augmentation techniques have been investigated, including cross domain adaptation \cite{bell2012transcription}, vocal tract length perturbation (VTLP) \cite{jaitly2013vocal,cui2015data,ko2015audio}, spectral distortion \cite{kanda2013elastic}, tempo perturbation  \cite{kanda2013elastic,ko2015audio}, stochastic feature mapping \cite{cui2015data}, speed perturbation \cite{ko2015audio} and end-to-end back translation \cite{hayashi2018back}. Various spectrogram deformations were applied in \cite{Park2019}.

In contrast, so far there is only limited research on data augmentation targeting disordered speech recognition. In \cite{christensen2013combining}, normal speech data was applied in the bottleneck feature extraction stage. In \cite{vachhani2018data}, data augmentation based on time- and tempo-stretching of normal speech was investigated. In \cite{jiao2018simulating}, a voice conversion framework was proposed to convert normal speech to disordered speech. In \cite{xiong2019phonetic}, phonetic analysis was conducted to compute the speaker-dependent scaling factors. These were then used in tempo adjustment to augment the limited disordered speech data. There are two issues associated with previous research. First, previous work in this direction provided piece-wise solutions, lacking a systematic comparison between augmentation approaches for disordered speech recognition. Second, the main focus of previous research has been on adjusting normal speech towards disordered speech, while investigation on augmentation using existing disordered speech data has been very limited. 

This paper presents a systematic investigation of different data augmentation techniques for disordered speech recognition. Augmented data was generated from two different sources: i) modification of healthy speech to disordered speech, and ii) perturbation of existing disordered speech. Three data augmentation techniques were used, including vocal tract length perturbation (VTLP), tempo perturbation and speed perturbation. Tempo perturbation stretches the length of the utterance but keeps the shape of spectral envelope unchanged \cite{verhelst1993overlap}, while speed perturbation resamples signals in time domain \cite{ko2015audio}. Speaker-level perturbation factors were obtained based on phonetic analysis described in \cite{xiong2019phonetic} and then used on normal speech. Global perturbation factors were used on disordered speech. Performance evaluation was conducted on a speaker independent DNN system implemented using an extended version of the Kaldi toolkit \cite{povey2011kaldi} for the Universal Access Speech (UASpeech) database. Learning hidden unit contributions (LHUC) \cite{swietojanski2016learning} based speaker adaptive training (SAT) was further applied to model the large variability among disordered speakers in both the original and augmented data.

The main contributions of this paper are summarized below. To the best of our knowledge, this is the first work to systematically investigate different data augmentation techniques for disordered speech recognition. Both normal and disordered speech were exploited in the augmentation process. Out of the three augmentation techniques, applying speed perturbation to both normal and disordered speech was found to give the best performance. The system using the best augmentation approach based on speed perturbation produced 2.92\% absolute (9.3\% relative) word error rate (WER) reduction over the baseline system without data augmentation. The final speaker adaptive system using the augmented data gave an overall WER of 26.37\% on the test set containing 16 UASpeech dysarthric speakers. As far as we know, this is the best performance reported so far on UASpeech. 
 
The rest of this paper is organized as follows. A range of data augmentation techniques are presented in section 2. Section 3 describes the baseline multi-speaker adaptively trained DNN system architecture. Section 4 presents experiments and results on the UASpeech database. The last section concludes and discusses possible future works.

\section{Data Augmentation}

We investigate three data augmentation techniques for disordered speech recognition. Both disordered speech and normal speech were used in the augmentation process. The former applied global perturbation factors while the latter applied speaker dependent perturbation factors based on phonetic analysis.

\subsection{VTLP Based Data Augmentation}
The shape of vocal tract varies from speaker to speaker \cite{wakita1973direct}. Vocal tract length normalization (VTLN) was proposed to counteract such variations \cite{lee1996speaker, eide1996parametric}. By applying VTLN in the reverse manner, vocal tract length perturbation (VTLP) adds variability to the speech data by simulating different vocal tract lengths \cite{jaitly2013vocal}. Given a time-domain audio segment $x(t)$, we denote the corresponding frequency-domain representation as $X(f)$, which is the Fourier transform of $x(t)$ . VTLP applies a perturbation factor $\alpha$ taken from a discrete set (e.g. $\{0.9, 1.1\}$) along the frequency axis of $X(f)$. The output ${Y}(f)$ is given as:
\begin{equation}
\setlength{\abovedisplayskip}{3pt}
\setlength{\belowdisplayskip}{3pt}
    Y(f)=X({\alpha}f)
\end{equation}
In this way, VTLP perturbs the spectral envelope of the audio segment while keeping the audio duration unchanged.

\subsection{Tempo Perturbation Based Data Augmentation}
Tempo perturbation stretches the duration of the audio signal $x(t)$ while leaving the shape of its spectral envelope untouched \cite{kanda2013elastic, ko2015audio}. This is achieved by first decomposing the time-domain audio segment $x(t)$ into short analysis blocks and then relocating these blocks along the time axis to construct the perturbed output $y(t)$ \cite{driedger2016review}. A well known algorithm in this area is waveform similarity overlap-add (WSOLA), which makes $y(t)$ share the maximal similarity with $x(t)$ by finding the optimal position of each analysis block iteratively \cite{verhelst1993overlap,driedger2016review}.

Suppose the time-domain audio signal $x(t)$ is decomposed into short analysis blocks $\tilde{x}_{m}(r)$. These blocks are equally spaced along the time axis by $H_a$ (the analysis hopsize). Given a perturbation factor $\alpha$, the synthesis blocks $\tilde{y}(r)$ are relocated along the time axis by $H_s$ (the synthesis hopsize) given as:
\begin{equation}
\setlength{\abovedisplayskip}{3pt}
\setlength{\belowdisplayskip}{3pt}
    H_s=\alpha\cdot{H_a}
\end{equation}
WSOLA takes an iterative approach to update the positions of analysis blocks. For example, the center of $\tilde{x}_m(r)$ is shifted by $\Delta_{m} \in [ -\Delta_{max}, \Delta_{max}]$ along the time axis, where the optimal value of $\Delta_{m}$ is obtained by maximizing the cross-correlation between $\tilde{x}_m(r)$ and $\tilde{x}_{m-1}(r)$. This ensures that the periodic structures of the adjusted analysis frame are optimally aligned with structures of the previously copied synthesis frame in the overlapping region when both frames use the synthesis hopsize $H_{s}$. The hann window function $w(r)$ is then applied on the adjusted analysis block to compute the synthesis block $\tilde{y}_m(r)$ \cite{verhelst1993overlap,driedger2016review}. After finishing all iterations, the synthesis frames are processed in order to reconstruct the actual time-scale modified output signal $y(t)$ in a similar manner as conventional OLA \cite{allen1977short}. In this way, $y(t)$ keeps the same spectral envelop shape as $x(t)$, but has a different length.

\subsection{Speed Perturbation Based Data Augmentation}
Speed perturbation resamples the audio signal in time domain \cite{ko2015audio}. Given an audio segment $x(t)$, a perturbation factor $\alpha$ is applied along the time axis and gives the output $y(t)$ as:
\begin{equation}
\setlength{\abovedisplayskip}{3pt}
\setlength{\belowdisplayskip}{3pt}
    y(t)=x({\alpha}t)
\end{equation}
In frequency domain, this is equivalent to the following change:
\begin{equation}
\setlength{\abovedisplayskip}{3pt}
\setlength{\belowdisplayskip}{3pt}
    X(f) \longrightarrow \frac{1}{\alpha}{X(\frac{1}{\alpha}f)}
\end{equation}
where $X(f)$ and $\frac{1}{\alpha}{X(\frac{1}{\alpha}f)}$ represent the Fourier transform of $x(t)$ and $y(t)$ respectively. In this way, speed perturbation leads to both change in the audio duration and perturbation in the spectral envelope \cite{ko2015audio, anden2014deep}. Table \ref{tab:aug_tech} summarizes the implement domain and effects of VTLP, tempo perturbation (denoted as   ``tempo'') and speed perturbation (denoted as ``speed'').

\begin{table}[H]
  \caption{Comparison of implement domain and effects of VTLP, tempo perturbation and speed perturbation. ``\cmark" indicates that change occurs after perturbation.}
  \label{tab:aug_tech}
  \centering
  \renewcommand\arraystretch{1} 
  \begin{tabular}{c|c|c|c}
\hline
\hline
  &{\textbf{VTLP}} & \textbf{tempo} & \textbf{speed}  \\ 
  \hline
Implement Domain   & $X(f)$               & $x(t)$               & $x(t)$             \\
Signal Duration    & no change & \cmark   & \cmark  \\
Spectral Envelope & \cmark   & no change & \cmark   \\ 
\hline
\hline
\end{tabular}
\end{table}

\subsection{Speaker Dependent Perturbation Factor Estimation}

Given the large mismatch in speaking rate between normal and disordered speakers and high variability among disordered speakers, speaker dependent perturbation factors were applied when modifying normal speech to disordered speech.The factors were obtained based on phonetic analysis described in \cite{xiong2019phonetic}. We performed force alignment using a GMM-HMM system based on the HTK toolkit \cite{young2006htk} to get frame-level phoneme alignments, and then calculated the average phoneme duration of each speaker (denoted as $l_{C_i}$ for control speaker ${C_i}$ and $l_{D_j}$ for dysarthric speaker ${D_j}$). For dysarthric speaker ${D_j}$, we took the average of $l_{C_i}$ (denoted as $l_{\overline{C}}$) as the reference to compute the speaker dependent perturbation factor $F_{D_j}$, given as: 

\begin{equation}
\setlength{\abovedisplayskip}{3pt}
\setlength{\belowdisplayskip}{3pt}
    F_{D_j}=\frac{l_{\overline{C}}}{l_{D_j}} 
\end{equation}
$F_{D_j}$ was then used as the perturbation factor when modifying normal speech to the speech of dysarthric speaker ${D_j}$.

\section{Baseline ASR System Description}

This section describes the baseline ASR system in terms of the hybrid DNN architecture and learning hidden unit contributions (LHUC) \cite{swietojanski2016learning} based speaker adaptive training.

\subsection{Hybrid Deep Neural Network Architecture}
\begin{figure*}[tp]
  \centering
  \includegraphics[width=\linewidth]{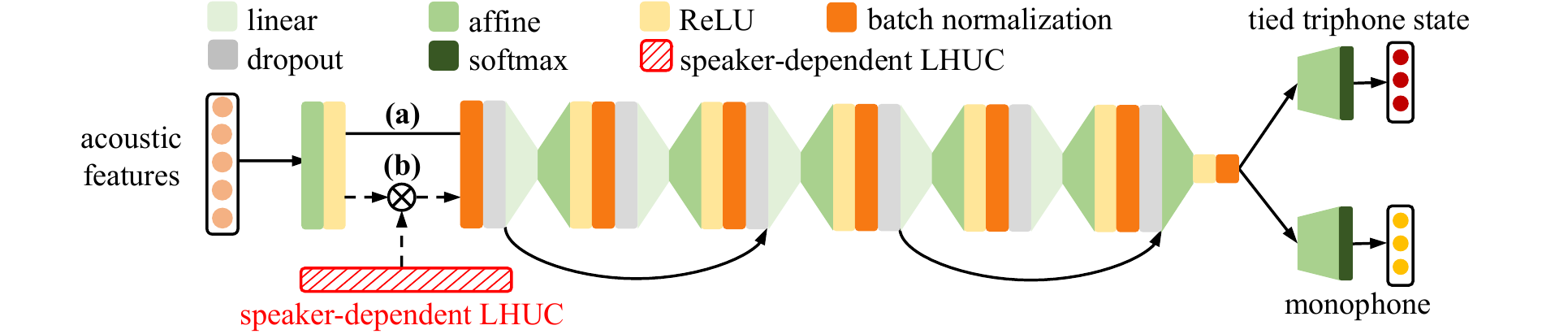}
  \caption{Architecture of the hybrid DNN system. The dashed line indicates the connecting component is an optional block. Retaining connection (a) leads to the speaker-independent baseline system, while retaining connection (b) leads to the system with LHUC SAT.}
  \label{fig:DNN_structure}
\end{figure*}

The structure of the hybrid deep neural network (DNN) acoustic model contains seven hidden layers. Each hidden layer contains a basic set of neural operations performed in sequence, i.e. affine transformation (in green), rectified linear unit (ReLU) activation (in yellow) and batch normalization (in orange). Apart from this, linear bottleneck projections (in light green) are applied to the inputs of the intermediate five hidden layers to reduce the number of network parameters. Dropout operations (in grey) are applied to the outputs of the first six hidden layers to prevent over-fitting. Softmax activation (in dark green) is used in the output layer. To speed up the training process and circumvent the vanishing gradient problem, two skip connections are used to connect the output of the first layer to the third layer and the output of the fourth layer to the sixth layer respectively. Figure \ref{fig:DNN_structure} illustrates the architecture of the hybrid DNN system.

Multi-task learning (MTL) \cite{caruana1997multitask} was used to train the hybrid DNN system shown in Figure \ref{fig:DNN_structure}. The labels for the two tasks are based on frame-level tied triphone states and monophone alignments respectively. The alignments were obtained from a GMM-HMM system implemented using the HTK toolkit \cite{young2006htk}. Incorporating frame-level monophone alignments in the labels reduces the risk of over-fitting to unreliable frame-level triphone states computed from disordered speech. The loss function of the mutli-task learning is as follows:

\begin{equation}
\setlength{\abovedisplayskip}{3pt}
\setlength{\belowdisplayskip}{3pt}
    \mathcal{L}_{MTL}={\lambda}{\cdot}\mathcal{L}_{tristate} +(1-\lambda){\cdot}\mathcal{L}_{mono}
\setlength{\abovedisplayskip}{3pt}
\setlength{\belowdisplayskip}{3pt}
\end{equation}
where $\mathcal{L}_{tristate}$ is the cross-entropy loss for the tied triphone state task, $\mathcal{L}_{mono}$ is the cross-entropy loss for the monophone task, and $0\leq\lambda\leq1$ is a tunable task weight parameter.

\subsection{LHUC Based Speaker Adaptive Training}
To model the large variability among disordered speakers, learning hidden unit contributions (LHUC) \cite{swietojanski2016learning} based speaker adaptive training (SAT) was used. Speaker-level LHUC scaling factors were applied to the ReLU activation output in the first layer, as shown by connection (b) in Figure \ref{fig:DNN_structure}. Supervised estimation of the LHUC factors was performed during the training stage, where the LHUC factors were updated once per mini-batch together with the network parameters. Unsupervised LHUC adaptation was performed during the test stage, where the LHUC factors were updated once per utterance.

\section{Experiments and Results}
\subsection{Task Description}
The UASpeech \cite{kim2008dysarthric} is an isolated word recognition task consisting of 16 dysarthric and 13 control speakers. The speech materials contain 155 common words and 300 uncommon words. There are 3 blocks per speaker, each containing all 155 common words and one third of the uncommon words. We treated block 1 and block 3 of all 29 speakers as training set, and block 2 of the 16 dysarthric speakers as test set. Silence stripping was performed using a GMM-HMM system to remove redundant silence in the recordings, as described in our previous work \cite{yu2018development}. For the baseline system without data augmentation, the training set contained 99195 utterances ($\sim$30.6 hours) and the test set contained 26520 utterances ($\sim$9 hours).

\subsection{Experiment Setup}
\begin{figure}[H]
  \centering
  \includegraphics[width=\linewidth]{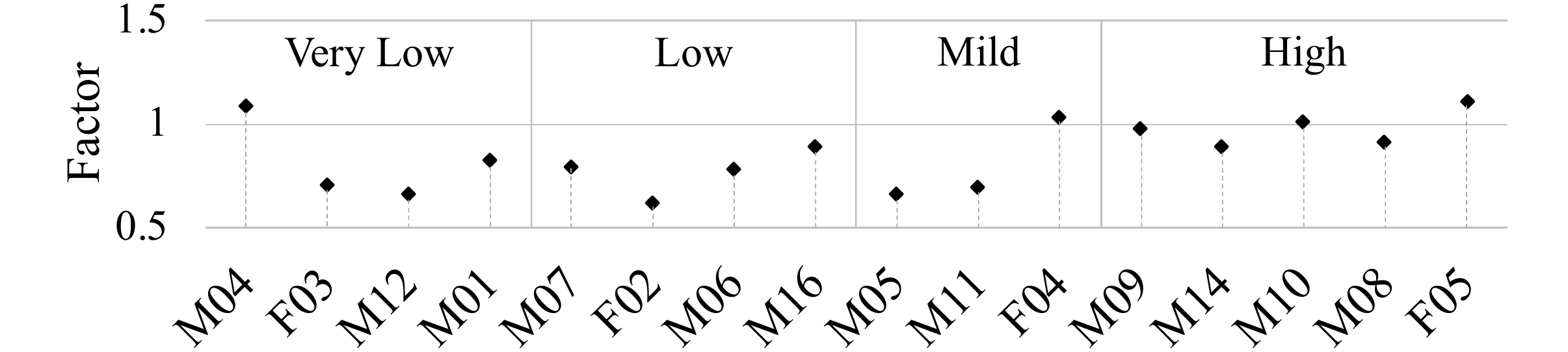}
  \caption{Speaker-level scaling factors for modifying normal speech to target disordered speech in the order of intelligibility \cite{christensen2012comparative, kim2008dysarthric}. A factor $F < 1$ slows down the normal speech while $F > 1$ speeds up the normal speech.}
  \label{fig:factor}
\end{figure}

We implemented the techniques discussed in section 2 to perform data augmentation on the training set and left the test set untouched. The HTK toolkit \cite{young2006htk} was used for VTLP. The tempo command of the Sox toolkit \cite{sox} was used for tempo perturbation, which is based on WSOLA. The speed command of Sox was used for speed perturbation. Following \cite{ko2015audio}, three sets of global perturbation factors were applied on the disordered speech: $\{0.9,1.1\}$, $\{0.9,0.95,1.05,1.1\}$ and $\{0.85,0.9,0.95,1.05, 1.1,1.15\}$. Speaker dependent perturbation factors were applied when modifying normal speech to target disorder speech, as discussed in section 2.4. The speaker-level scaling values are shown in Figure \ref{fig:factor}.

In our experiments, the hybrid DNN acoustic model was implemented using the Kaldi toolkit \cite{povey2011kaldi}. A 9-frame context window was used in the system. The inputs were 80-dimension filter bank (FBK) + $\Delta$ features. The first six hidden layers contain 2000 neurons each, while the dimension of the linear bottleneck projections is 200 and dropout rate is 20\%. The seventh hidden layer contains 100 neurons. The system was trained by back-propagation based on RMSProp optimizer. For multi-task learning, the task weight parameter $\lambda$ was set as 0.5, giving the same weight to both tied triphone state and monphone task. Following \cite{christensen2012comparative}, a uniform language model was used in decoding.

\subsection{Performance of Data Augmentation }

\begin{table*}[tp]
  \caption{Performance on the 16 UASpeech dysarthric speakers of different data augmentation approaches applied on the training set. Here ``CTL'' / ``DYS''  stands for normal / disordered speech.  ``tempo''  / ``speed'' stands for tempo / speed perturbation. ``2x'', ``4x'' and ``6x'' refer to the amount of augmented data. ``Very low'', ``Low'', ``Mid'' and ``High'' refer to intelligibility of the DYS groups.  }
  \label{tab:result}
  \centering
  \renewcommand\arraystretch{0.95} 
\begin{tabular}{c|c|c|c|c|c|cccc|c}
\hline
\hline
\multirow{2}{*}{Sys. ID}& \multicolumn{3}{c|}{Data Augmentation}   & \multirow{2}{*}{\# Hours} & LHUC & \multicolumn{5}{c}{WER\%}  \\
\cline{2-2} \cline{3-3} \cline{4-4}  \cline{7-7} \cline{8-8}  \cline{9-9} \cline{10-10} \cline{11-11}
    ~   & Method               & CTL & DYS & ~ & SAT & Very low & Low  & Mid & High   & Overall \\
\hline
\hline
1      & \multicolumn{3}{c|}{NA} & 30.6 & \xmark  & 69.82  & 32.61        & 24.53    & 10.40  & 31.45   \\
2       & \multicolumn{3}{c|}{NA} & 30.6 & \cmark        & 64.39  & 29.88        & 20.27    & 8.95  & 28.29   \\
\hline
\hline
3      & VTLP                 & 1x  &  - & 48.0  & \xmark      & 68.68  & 31.84        & 22.71    & \textbf{9.48}  & 30.35   \\
4      & Tempo                & 1x  & -  & 52.2   & \xmark      & 70.71  & 32.78        & 25.12    & 10.32 & 31.77   \\
5      & Speed                & 1x  & - & 52.2    & \xmark      & \textbf{67.52}  & \textbf{31.55}  & \textbf{21.96}    & 9.57  & \textbf{29.92}   \\
\hline
\hline
6      & VTLP                 &  -   & 2x & 65.5  & \xmark          & 69.98  & 30.08        & 21.39    & 9.65  & 29.97   \\
7      & Tempo                &  -   & 2x & 65.9  & \xmark          & 69.32  & 31.75        & 23.94    & 10.07 & 30.90    \\
8      & Speed                &  -   & 2x & 65.9 & \xmark          & 68.43  & 29.60         & 21.37    & 10.44 & \textbf{29.79}   \\
\hline
9      & Speed                &  -   & 4x & 100.9 & \xmark          & 67.20   & \textbf{29.86}        & 21.45    & \textbf{10.04} & \textbf{29.47}   \\
10     & Speed                &  -   & 6x & 136.7 & \xmark          & \textbf{67.15}  & 30.07        & \textbf{21.25}    & 10.17 & 29.52   \\
\hline
\hline
11     & Speed                & 2x  & 2x & 130.1  & \xmark          & 66.45  & 28.95        & 20.37    & 9.62  & 28.73   \\
12     & Speed                & 4x  & 4x & 207.5 & \xmark          & 66.26  & 28.60         & 19.90     & 9.68  & \textbf{28.53}   \\
\hline
13     & Speed                & 2x   & 2x & 130.1  & \cmark       & 62.50  & \textbf{27.26}        & 18.41    & 8.04  & 26.55   \\
14     & Speed                & 4x  & 4x & 207.5 & \cmark        & \textbf{62.44}  & 27.55        & \textbf{17.35}    & \textbf{7.93}  & \textbf{26.37}   \\
\hline
\hline    
\end{tabular}
\end{table*}

Table \ref{tab:result} shows the performance of different data augmentation approaches applied on the training set. ``2x'', ``4x'' and ``6x'' refer to the amount of augmented data. Several trends can be observed. \textbf{1)} Sys.3-5 show that for augmentation using normal speech, speed perturbation gives the best overall performance with an absolute WER reduction of 1.53\% over the baseline (Sys.1). \textbf{2)} Sys.6-8 demonstrate that for augmentation using disordered speech, speed perturbation also gives the best overall performance. Therefore, we proceed with speed perturbation to generate more augmented data in the remaining experiments. \textbf{3)} Sys.8-10 indicate that perturbing the disordered speech by $\{0.9,0.95,1.05,1.1\}$(4x) works better than $\{0.9,1.1\}$(2x) or $\{0.85,0.9,0.95,1.05,1.1,1.15\}$(6x). \textbf{4)} Sys.8-9,11-12 show that augmentation using both normal speech (with speaker dependent factors) and disordered speech (with global factors) can further improve the performance by an absolute WER reduction of 0.94\% (Sys.12 over Sys.9), while increasing the amount of augmented data produces only marginal improvement (Sys.12 over Sys.11). The best speaker-independent system with data augmentation (Sys.12) produces \textbf{2.92\%} absolute (9.3\% relative) WER reduction over the speaker-independent baseline system without data augmentation (Sys.1).

\subsection{Peformance of LHUC SAT}

LHUC based speaker adaptive training was further applied to model the variability among disordered speakers in both the original and augmented data (see Table \ref{tab:result}, Sys.2,13-14). The augmented data was assigned to individual target dysarthric speakers for the speaker-level LHUC adaptation. This leads to a further 2.16\% absolute WER reduction (Sys.14 over Sys.12), indicating that LHUC SAT contributes to modelling the large variability among disordered speakers. The final speaker adaptive system using the best augmentation approach gives an overall WER of 26.37\% on the test set containing 16 UASpeech dysarthric speakers (see Table \ref{tab:result}, Sys.14). To the best of our knowledge, this corresponds to an absolute WER reduction of 1.51\% over the best published system on UASpeech, as shown in Table \ref{tab:compare}.

\begin{table}[H]
  \caption{A comparison between published systems on UASpeech and our system. Here ``DA'' refers to data augmentation.}
  \label{tab:compare}
  \centering
   \renewcommand\arraystretch{1}
\begin{tabular}{cc}
\toprule  
    Systems   &  WER\%   \\
\midrule
    Sheffield-2013 Cross domain augmentation \cite{christensen2013combining}    & 37.50 \\
    Sheffield-2015 Speaker adaptive training  \cite{sehgal2015model}    & 34.80  \\
    CUHK-2018 DNN System Combination \cite{yu2018development}         & 30.60  \\
    Sheffield-2019 Kaldi TDNN + DA \cite{xiong2019phonetic}    & 27.88  \\
    \textbf{Speed perturb + LHUC SAT (Table 2, Sys.14)}  & \textbf{26.37} \\ 
\bottomrule
\end{tabular}
\end{table}

\section{Conclusions}
This paper presents a systematic investigation of different data augmentation techniques for disordered speech recognition. It suggests that speed-perturbation based augmentation produces the largest improvement in system performance despite the huge mismatch between normal and disordered speech. Future research will focus on more powerful data augmentation techniques to cover other features of disordered speech, such as articulation imprecision, reduced intensity and disfluency.

\section{Acknowledgement}
This research is supported by Hong Kong Research Grants Council General Research Fund No. 14200218, Theme Based Research Scheme T45-407/19N and Shun Hing Institute of Advanced Engineering Project No. MMT-p1-19. The authors would like to thank Dr. Feifei Xiong for fruitful discussions. 

\bibliographystyle{IEEEtran}

\bibliography{mybib}

\end{document}